\documentclass[prl,twocolumn,showpacs,superscriptaddress,floatfix]
{revtex4}
\usepackage{graphics,epsfig,amsfonts,amssymb,amsmath,ulem,color}
\begin{document}
\title{Nematicity at the Hund's metal crossover in iron superconductors}
\author{L. Fanfarillo$^*$}
\affiliation{CNR-IOM and International School for Advanced Studies (SISSA), Via Bonomea 265, I-34136, Trieste, Italy}
\author{G. Giovannetti}
\affiliation{CNR-IOM and International School for Advanced Studies (SISSA), Via Bonomea 265, I-34136, Trieste, Italy}
\author{M. Capone}
\affiliation{CNR-IOM and International School for Advanced Studies (SISSA), Via Bonomea 265, I-34136, Trieste, Italy}
\author{E. Bascones$^*$}
\affiliation{Instituto de Ciencia de Materiales de Madrid, 
ICMM-CSIC, Cantoblanco, E-28049 Madrid (Spain).}
\email{laura.fanfarillo@sissa.it}
\email{leni@icmm.csic.es}
\date{\today}
\begin{abstract} 
{ The theoretical understanding of the nematic state of iron-based
superconductors and especially of FeSe is still a puzzling problem. Although a
number of experiments calls for a prominent role of local correlations and place
iron superconductors at the entrance of a Hund metal state, the effect of the
electronic correlations on the nematic state 
has been theoretically poorly investigated. In this work we study the nematic
phase of iron superconductors accounting for local correlations, including the
effect of the Hund's coupling. We show that Hund's physics strongly affects the
nematic properties of the system. It severely constraints the precise nature of
the feasible orbital-ordered state and
induces  a differentiation in the effective masses of the $zx/yz$ orbitals in
the nematic phase. The latter effect leads to distinctive signatures in different
experimental probes, 
so far overlooked in the interpretation of experiments.
As notable examples the splittings between $zx$ and $yz$
bands at $\Gamma$ and M points are modified, with important consequences for
ARPES 
measurements.
}
\end{abstract} 


\maketitle
\section{Introduction} 
The nematic state is one of the most debated issues in iron-based
superconductors (IBS) \cite{fisher_reviewnematic2011, fernandes_natphys,
gallais_reviewnematic2016, bohmer_reviewnematic2016}.
Perturbative  approaches which attribute the ordered phases to Fermi- surface
instabilities have succeeded in explaining many properties of IBS, including the
occurrence of a nematic phase. Within this framework, the Ising spin-nematic
model \cite{kivelson_nematic, sachdev_nematic, fernandes_prb2012,
fernandes_natphys,fanfarillo_prb2015b}, interpreted  the nematic state as a
precursor of the $(\pi,0)$ stripe order.

On the other hand, many theoretical studies \cite{werner_prl2008,
haule_njp2009, demedici_prl2009, ishida_prb2010, liebsch_prb2010,
zpyin-natmaterials2011, werner2012, demedici_prl2011, yu_prb2012,
nosotras_prb2012-2, lanata_prb2013, demedici_prl2014, fanfarillo_prb2015,
nosotras_review2016} have highlighted the relevance of electron correlations
driven by local interactions, and in particular by the Hund's coupling, to
explain the bad metallic behavior of the normal state of IBS. A number of
experimental investigations support indeed this scenario \cite{yi_natcomm2015,
aronson_prb2015, demedici_prl2014, meingast_prl2013, Hardy2016}. 

Several theoretical investigations agree on a central role of the Hund's
coupling and link the phenomenology of IBS to a crossover between a normal metal
and the so-called Hund's metal which occurs when the interactions are increased.
In particular, the IBS lie close to this crossover, which is characterized by a
suppression of the overall coherence of conduction electrons and it is
accompanied by a pronounced differentiation of the correlation strengts between
different orbitals \cite{demedici_prl2014, fanfarillo_prb2015}. As a matter of
fact the system appears in this region particularly sensitive to any kind of
perturbation. Therefore accounting for local correlations is crucial for any realistic
theoretical study of the nematic phase of IBS.

\begin{figure*}
\leavevmode
\includegraphics[clip,width=0.99\textwidth]{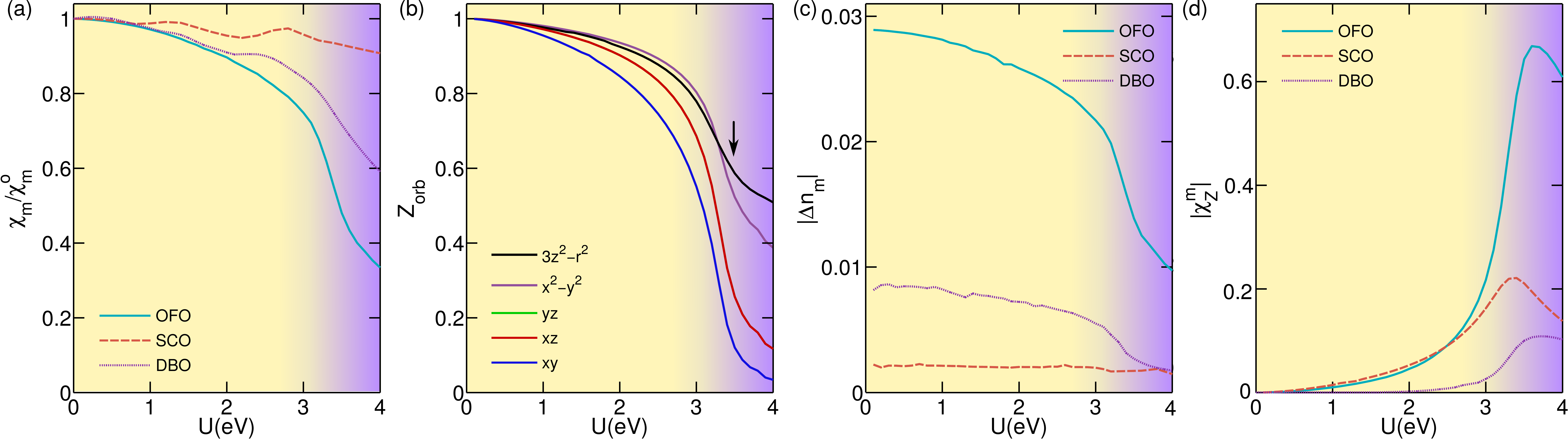} \caption{(Color 
online)(a) Orbital nematic susceptibilities $\chi_m$(U) normalized to their 
respective $U=0$ value, for $J_H/U=0.20$ and electronic filling $n=6$ 
\cite{ref_plane}. The OFO is strongly suppressed by interactions while the SCO 
is only weakly affected. (b) Orbital-dependent quasiparticle weight $Z_{\gamma}$ 
as a function of $U$ in the absence of a nematic perturbation. The sharp drop 
around $U\sim3.2$ eV signals the entrance into the strongly correlated Hund's 
metal region where the atoms become spin polarized. The shading colors in the 
panels differentiate the weakly correlated and the Hund's metal regime. The 
arrow signals the interaction value $U=3.5$ eV at which the orbital-dependent 
mass enhancements equal those observed in ARPES and Quantum Oscillations 
experiments in FeSe \cite{maletz_prb2014, coldea_prb2015, ishizaka_prb2015}. (c) 
Occupation imbalance between $zx$ and $yz$ orbitals for the three orders taking  
$2 h_m=75 meV$. The larger is the imbalance produced by an order, the stronger 
is the $\chi_m$(U) suppression of this order by interactions. (d) Absolute value 
of the $Z$-response functions $\chi^{m}_Z(U)$. 
} 
\label{fig:Fig1} 
\end{figure*}

The nematic state of FeSe deserves a particular interest. Even if sizable spin
fluctuations have been detected \cite{wang_natmat,rahn_prb,lo_arxiv2016}, the
lack of magnetic order in this material casts doubts on a magnetic origin of
nematicity challenging the Ising spin-nematic picture. This has reinvigorated
approaches based on orbital ordering. Orbital order models have mostly focused
on states with occupation imbalance between $zx$ and $yz$ orbitals
\cite{phillips_prb2009, kruger_prb2009, leeyinku_prl2009,onari_prl2012}. However
calculations on realistic IBS tight-binding models using FLEX
\cite{onari_prb2012} and Hartree-Fock methods \cite{nosotras_prl2010-1,
nosotras_prl2010-2} do not find a similar instability. 
In fact Hund's coupling favors an even population of the orbitals in order to maximize
the spin.

More recently, it has been proposed that FeSe can host orbital ordering with
$n_{zx}\neq n_{yz}$ mediated by spin fluctuations because of the relatively
small ratio between Hund's coupling $J_H$ and intraorbital interactions $U$
\cite{kontani_prx2016}. This argument however neglects that FeSe is one of the
most correlated IBS \cite{nosotras_review2016, liebsch_prb2010, lanata_prb2013}
as testified by the the recent observation of the Hubbard
bands\cite{borisenkohubbard,coldeahubbard}, the large mass enhancements
\cite{maletz_prb2014, coldea_prb2015, ishizaka_prb2015} and the
orbital-dependent coherence-incoherence crossover \cite{yi_natcomm2015} ,
features typical of Hund's metals. The value of $U$, larger in FeSe than in
other IBS \cite{miyake_jpsj2010} makes Hund's physics relevant in FeSe despite a
smaller $J_H/U$ ratio and requires including the correlations beyond the
approximation used in \cite{kontani_prx2016}.  

Besides the onsite ferro-orbital ordering (OFO) discussed above, other orbital
orders have been proposed to explain the nematic state of FeSe. In particular:
(i) An orbital polarization with sign reversal between electron (e) and hole (h)
pockets (``sign-changing bond order'', SCO) emerges from parquet renormalization
group (RG) calculations when the Fermi energy is small, as in FeSe
\cite{chubukov2016}; (ii)  A nematic d-wave bond order (DBO) \cite{su_jpcm2015}
stabilized by a large intersite repulsion \cite{hu_dwave2016}, was proposed to
explain early ARPES experiments in FeSe \cite{ishizaka_prb2014,
ishizaka_prb2015, coldea_prb2015, ding_prb2015, shen_FeSenematic2015},
which at present are controversial \cite{watson2016, brouet2016, borisenko2016}. 

The predictions for the stability and the spectroscopic signatures of the
orbital-ordered states discussed so far are based on techniques unable to
address the effect of local correlations and could change once Hund's physics is
taken into account. In this work we study the impact of short-range correlations
on the anisotropic properties of IBS in the nematic state including the
possibility that local correlations induce a phase transition to an
orbital-ordered state. We use slave-spin calculations to study whether Hund's
physics coexist or compete with three different orbital orders: onsite
ferro-orbital order OFO, SCO  which changes sign between the e-/h-pockets and
d-wave bond nematic order DBO. We find that: (i) Local correlations alone are
not able to stabilize a nematic orbital phase but constraints the nature of the
possible orbital orders. In particular Hund's coupling strongly suppresses those
orders which produce larger uneven occupation of $zx$ and $yz$ orbitals, as the
OFO, but it has a minor effect on the orders for which the occupation imbalance
is small, as the SCO; (ii) The orbital decoupling induced by interactions at the
Hund's metal crossover favors a differentiation of the $zx$/$yz$ orbitals masses
in the nematic phase; (iii) Interactions modify the orbital splittings between
$yz$ and $zx$ at the $\Gamma$ and M symmetry points with strong implications in
the interpretation of ARPES experiments.

\section{The model} 
We consider a five-orbital Hubbard-Kanamori 
Hamiltonian, widely used to study IBS
\begin{eqnarray}
\nonumber
& H &  = \sum_{i,j,\gamma,\beta,\sigma}t^{\gamma,\beta}_{i,j}c^\dagger_{i,\gamma,\sigma}c_{j,\beta,\sigma}+h.c. 
+ U\sum_{j,\gamma}n_{j,\gamma,\uparrow}n_{j,\gamma,\downarrow}
\\ \nonumber & +&  (U'-\frac{J_H}{2})\sum_{j,\gamma>\beta,\sigma,\tilde{\sigma}}n_{j,\gamma,\sigma}n_{j,\beta,\tilde{\sigma}}
-2J_H\sum_{j,\gamma >\beta}\vec{S}_{j,\gamma}\vec{S}_{j,\beta}
\\
& + &  J'\sum_{j,\gamma\neq
  \beta}c^\dagger_{j,\gamma,\uparrow}c^\dagger_{j,\gamma,\downarrow}c_{j,\beta,\downarrow}c_{j,\beta,\uparrow}
+ \sum_{j,\gamma,\sigma}\epsilon_\gamma n_{j,\gamma,\sigma} \,.
\label{eq:hamiltonian}
\end{eqnarray}
 $i,j$ label the Fe sites in the 1-Fe unit cell, $\sigma$ the spin and 
$\gamma, \beta$ the five Fe d-orbitals $yz$, $zx$, $xy$, $3z^2-r^2$ and 
$x^2-y^2$, with $x$ and $y$ axis along the Fe-Fe bonds. 
For the tight-binding Hamiltonian we refer to the parameters used in  
\cite{demedici_prl2014} for FeSe. $U$ and $U'$ are the 
intra- and inter-orbital interactions, $J_H$ is the Hund's coupling and 
$J'$ the pair-hopping and we assume $U'=U-2J_H$.
Except otherwise indicated we use $J_H=0.2 U$ and
electronic filling $n=6$ as in undoped IBS. 
We consider only the density-density part of the interaction, that we treat with
the single-site slave spin mean-field \cite{demedici_prb2005,
demedici_prb2010,fanfarillo_prb2015,demedicicapone2016}. 
This method accounts for the effect of local correlations, and gives the
correlated value of the orbital dependent fillings $n_\gamma$ and mass
enhancements which produces the band narrowing observed experimentally. At the
single-site level used here, the mass enhancements  equal the inverse of the
quasiparticle spectral weights $Z_\gamma$ discussed below. Collective modes such
as long-range spin-fluctuations are not included in the present description.

To study the interplay between local correlations and orbital 
order we add a small nematic perturbation to the original Hamiltonian 
Eq.~$(\ref{eq:hamiltonian})$ and analyze the linear response of the system. 
Perturbations in both the $A_{1g} \sim (n_{xz}({\bf k}) + n_{yz}({\bf k}))$ and 
$B_{1g} \sim (n_{xz}({\bf k}) - n_{yz}({\bf k}))$ channels are studied
\begin{equation}
\delta H^m_{A_{1g}/B_{1g}}  = \sum_{\bf k} (n_{xz}({\bf k}) \pm n_{yz}({\bf k})) f_m({\bf k}) h_m
\end{equation}
$m$ defines the orbital order under consideration, $f_m({\bf k})$ describes the
modulation of the perturbation in momentum space and $h_m$ is the perturbing
field. $\Delta_m=-\langle \sum_{\bf k} (n_{xz}({\bf k}) \pm n_{yz}({\bf k})) f_m
({\bf k})\rangle$ is the order parameter associated to each perturbation and 
$ \chi_m= \delta \Delta_m / \delta h_m $ is the relative susceptibility. 
A diverging susceptibility signals a second order phase transition while the
suppression of the susceptibility indicates competition between local
correlations and nematicity. 

Specifically we study:
 
(a) The $B_{1g}$ OFO perturbation which lifts the degeneracy of the $xz$/$yz$
onsite energy 
$$h_{OFO} = \delta \epsilon \quad \quad f_{OFO}({\bf k})= 1$$ 

$\Delta_{OFO} = \langle\sum_{\bf k} (n_{yz}({\bf k})-n_{xz}({\bf k}))f_{OFO}({\bf k})\rangle$ 
is the occupation imbalance between $xz$ and $yz$ orbitals 
\cite{phillips_prb2009, kruger_prb2009, leeyinku_prl2009}; 

(b) The $B_{1g}$ SCO perturbation given by 
$$h_{SCO} = \delta t^{\prime} \quad \quad  f_{SCO}({\bf k})= \cos k_x \cos k_y$$ 
that breaks the degeneracy of the 
intraorbital $zx$ and $yz$ second neighbor hoppings. The order parameter
$\Delta_{SCO}= \langle\sum_{\bf k} (n_{yz}({\bf k})-n_{xz}({\bf k}))f_{SCO}({\bf
k})\rangle $, that changes sign between the e- and h-pockets, is
compatible with predictions from RG calculations \cite{chubukov2016}; 

(c)The $A_{1g}$ DBO 
perturbation
$$h_{DBO} = \delta t \quad \quad  f_{DBO}({\bf k})= (\cos kx - \cos ky)/2$$ 
breaks the orbital degeneracy of the intraorbital $zx$ and $yz$ nearest neighbor 
hoppings. Its order parameter $\Delta_{DBO}= -\langle \sum_{\bf k} (n_{xz}({\bf 
k}) + n_{yz}({\bf k}))f_{DBO}({\bf k})\rangle $ was proposed in connection to 
ARPES experiments \cite{su_jpcm2015,ding_prb2015,shen_FeSenematic2015,hu_dwave2016}.

Below we refer specifically to the case of FeSe. Nevertheless our discussion as
a function of $U$ and main conclusions are valid for other IBS. 

\section{Interplay between Hund's coupling and orbital orders}

\begin{figure}
\leavevmode
\includegraphics[clip,width=0.45\textwidth]{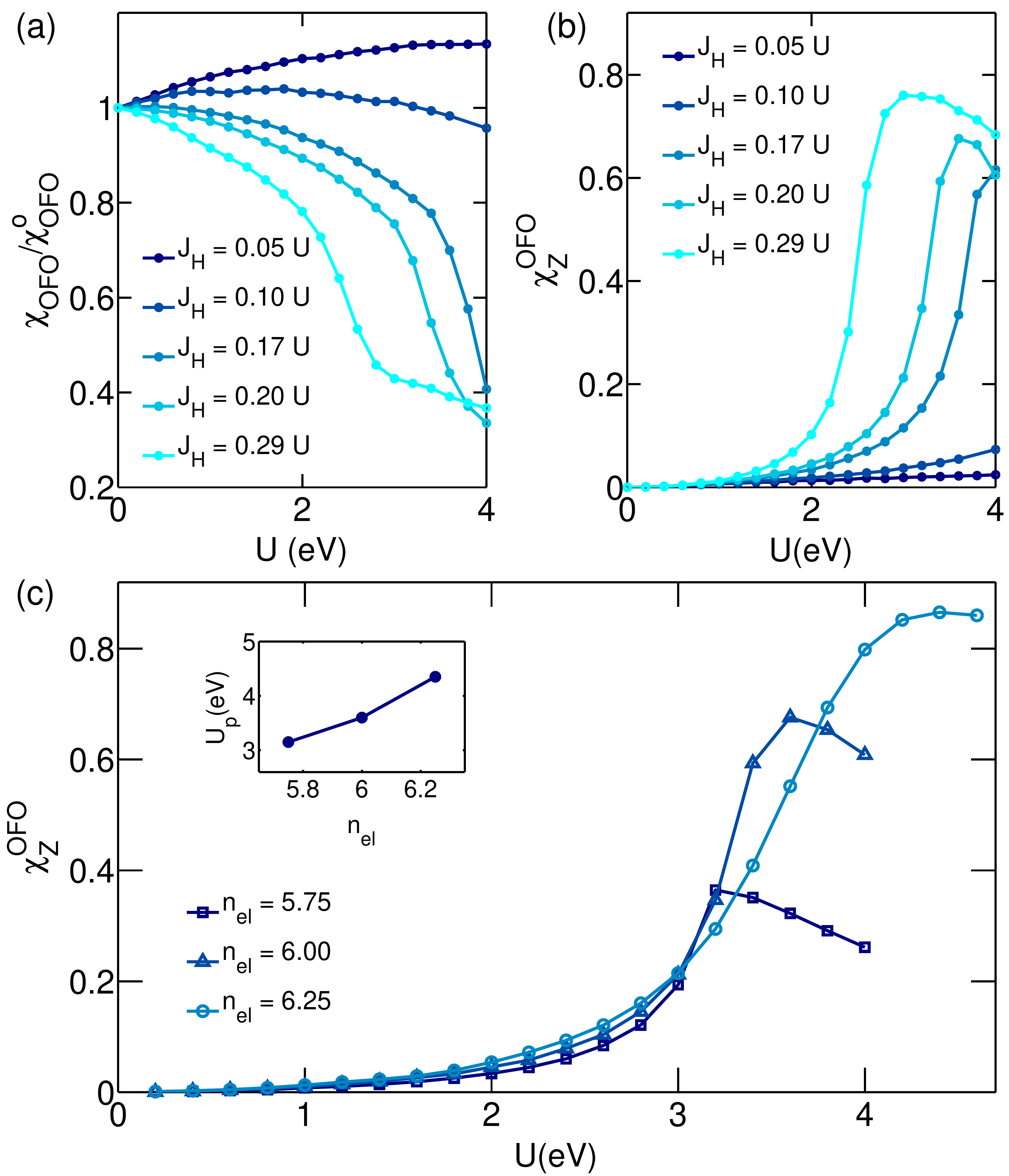}
\caption{ (Color online) (a) $\chi_{OFO}(U)$ susceptibility, normalized to its
$U=0$ value, for $n=6$ and different values of $J_H/U$. The suppression of the
order is prominent and appears at smaller $U$ for larger $J_H/U$. At small
$J_H/U$ the order is not suppressed. (b) $\chi^{OFO}_Z(U)$ for $n=6$ and
different values of $J_H/U$. (c) $\chi^{OFO}_Z(U)$ for $J_H/U=0.20$ and several
$n$. 
In the inset: Interaction $U_p$ at which $\chi^{OFO}_Z$ peaks as a function of
$n$. The peak in $\chi^{OFO}_Z$ moves towards smaller $U$ with increasing
$J_H/U$ and decreasing $n$ following the same dependence than the crossover
between the weakly correlated and the Hund metal. 
The evolution of $\chi^{OFO}_Z$ shown in panels (b)-(c) is 
representative also for the DBO and SCO case.} 
\label{fig:Fig2} 
\vspace{-0.4cm}
\end{figure}

Fig.~\ref{fig:Fig1}(a) shows the nematic susceptibilities
$\chi_m$ as a function of $U$, normalized to their non-interacting value $\chi^0_m$. 
The OFO susceptibility is strongly suppressed by
interactions. This effect closely tracks the strengthening of local
correlations, measured by the orbital-dependent quasiparticle weights 
$Z_{\gamma}$ which fall at the same interaction values, see
Fig.~\ref{fig:Fig1}(b). Correspondingly the electronic mass, not shown
explicitly, are strongly enhanced at these interactions. The sharp drop of
$Z_{\gamma}$ evidences the crossover between two metallic states: a low-spin
weakly correlated state and a high-spin strongly correlated metallic state
\cite{yu_prb2012, fanfarillo_prb2015, nosotras_review2016}, the Hund metal. 
The suppression of the OFO order is prominent at the entrance of the Hund's
metal region, as also seen in Fig.~\ref{fig:Fig2}(a):
With increasing $J_H/U$ the suppression of the order becomes sharper and happens
at smaller values of $U$, following the evolution of the Hund's metal crossover
with $J_H/U$ \cite{demedici_prl2011, yu_prb2012, fanfarillo_prb2015,
nosotras_review2016}. In contrast with $\chi_{OFO}$, $\chi_{SCO}$ depends weakly
on $U$, while $\chi_{DBO}$ is suppressed, but not so strongly as $\chi_{OFO}$ (Fig.~\ref{fig:Fig1}(a)). 

None of the susceptibilities diverge as a function of the interactions. This
indicates that the local correlations do not produce by itself the nematic
transition. Nevertheless, the different response of the system to each of the
orders at the Hund metal crossover found above  implies that Hund's physics
strongly constraints the nature of the nematic state.

Quantum oscillation and ARPES experiments on FeSe give mass renormalization
factors $m_{zx,yz}\sim 3$ and $m_{xy}\sim 6-9$  \cite{maletz_prb2014,
coldea_prb2015, ishizaka_prb2015}. From the values in Fig.~\ref{fig:Fig1} (b),
we estimate $U\sim 3.5$ eV for FeSe. These mass enhancements  place FeSe very
close to the Hund metal crossover. For this value of the interaction OFO order
is strongly suppressed what practically rules out the possible presence this
order. Conversely, local correlations barely affect the SCO making this order a
good candidate to explain the nematic state of FeSe.

The different response of the system to the  different orders can be traced back
to the different occupation imbalance between $zx$ and $yz$ orbitals, which is
large in the OFO case, intermediate for the DBO and very small for the SCO
perturbation, see Fig.~\ref{fig:Fig1}(c). Hund's coupling spreads the electrons
among the orbitals and therefore strongly suppresses the orders implying a large
imbalance. Unequal orbital occupation is suppressed by Hund's coupling because
it reduces the spin polarization and increases the intraorbital double occupancy
which is energetically unfavorable (see Eq.~(\ref{eq:hamiltonian})). Calculations
with a different SCO perturbation \cite{hopping_note} but with similarly small
occupation imbalance give consistent results regarding the weak effect of
interactions and support this interpretation.  

The differences in the magnitude of the charge imbalance depend on the band
reorganization produced by the orbital order close to the Fermi level. The
contributions of the e- and h-pockets to the occupation imbalance add in the OFO
case, at $U=0$ while they partially cancel out in the SCO perturbation.
 
In the following, we consider that a nematic order is present, assuming that it
has been stabilized by degrees of freedom not included here, and analyze the
effect of local correlations on the anisotropic properties. For the sake of
generality we do not restrict the type of orbital order and consider the three
kinds discussed above.   

\begin{figure*}
\leavevmode
\includegraphics[clip,width=0.99\textwidth]{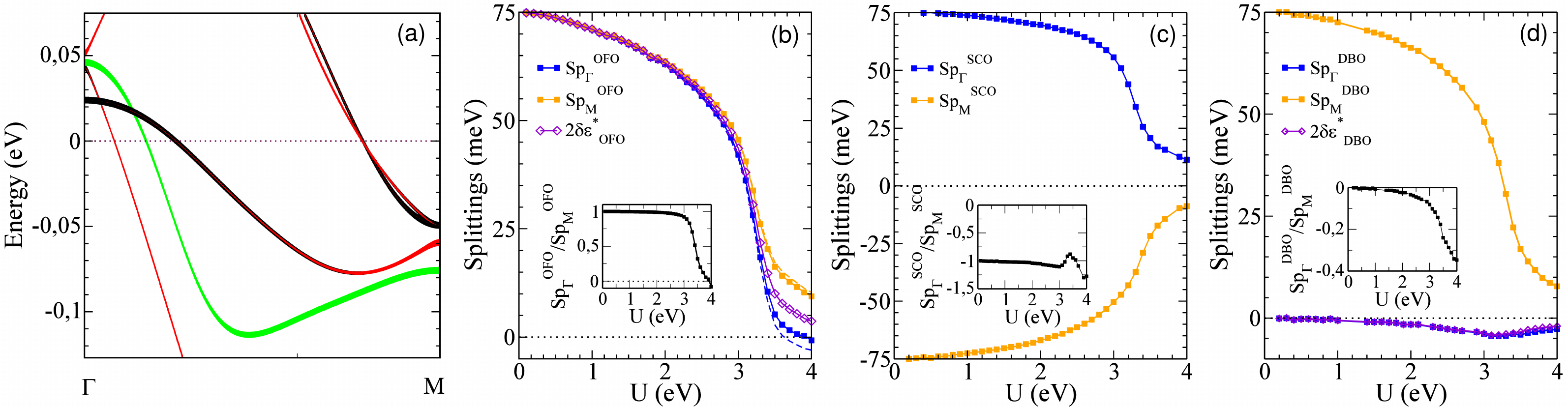} 
\caption{ (Color online) (a) FeSe band structure in the 2Fe Brillouin zone close to
the Fermi level along $\Gamma-M$ for an OFO
perturbation $2\delta\epsilon =75$ meV and $U=3.5$ eV ($zx$
is indicated in red, $yz$ in green and $xy$, $x^2-y^2$ and $3z^2-r^2$ in black).
(b)-(d) Splittings at $\Gamma$ and M, $Sp^m_\Gamma$ (blue) and $Sp^m_M$ (orange)
as a function of $U$ for the (b) OFO, (c) SCO and (d) DBO perturbations
corresponding to $2h_m=75 meV$. The renormalized onsite energy splitting for the
OFO case (b) $2 \delta \epsilon^*_{OFO} =2 \delta \epsilon-\delta \lambda^{OFO}$
and the DBO one (d) $2 \delta \epsilon^*_{DBO} = -\delta \lambda^{DBO}$ are shown
in purple\cite{ref_OFO}. In (c) $2 \delta \epsilon^*_{SCO}$ is smaller than 3 meV and not
shown. Solid lines are guides to the eye. Dashed lines in (b) give the value of
the splittings calculated with the approximation in Eq.
(\ref{eq:approxsplitting}).   With interactions 
the relative ratio of the splittings, shown in the insets, changes. 
The different value of $Sp^{OFO}_\Gamma$ and $Sp^{OFO}_M$ in (b) is due to the
different value of $Z_{zx}$ and $Z_{yz}$. On the other hand the finite value of
$Sp^{DBO}_\Gamma$ in (d) is primarily accounted for  by the change in the onsite
energies $2 \delta \epsilon^*_{DBO}$. Spin-orbit interaction, not included here,
will further modify the splittings at $\Gamma$ \cite{ref_plane}.}
\label{fig:Fig3} 
\end{figure*}  

{\section{Anisotropic orbital differentiation at the Hund's crossover.}}

In the tetragonal state the  $zx$ and $yz$ orbitals are degenerate and their
quasiparticle weights are equal. A nematic perturbation lifts this degeneracy
and leads to different $Z$'s, inducing a further source of anisotropy in the
observables. A different degree of correlations in $zx$ and $yz$ orbitals was
early suggested in Hartree-Fock calculations in the $(\pi,0)$ magnetic
state\cite{nosotras_prb2012-2}. However, this effect has been overlooked in the
theoretical studies of the nematic state and in the interpretation of the
experiments. To analyze this anisotropy we consider a generalized linear
response function
\begin{equation}
\chi^m_Z(U)=\frac{\delta(Z_{zx}-Z_{yz})}{\delta h_m}.
\end{equation}
For each of the three orders, $\chi^m_Z(U)$ shows a clear peak
at the Hund's metal crossover, see Fig.~\ref{fig:Fig1} (d). 
In this region the quasiparticle weight is more sensitive to small perturbations.
The spin polarization favors the decoupling of the orbitals as the effective
interaction between the electrons in different orbitals is reduced when their
spins are parallel \cite{demedici_prl2014, fanfarillo_prb2015}. 

The peak position has the same dependence on the  ratio $J_H/U$ and on the
filling $n$ as the one previously reported for the Hund's metal
crossover\cite{werner_prl2008,demedici_prl2011,demedici_prb2011,
yu_prb2012,fanfarillo_prb2015,nosotras_review2016}. 
The dependence of the peak position can be seen in Fig. \ref{fig:Fig2} (b) and
(c) for the OFO case but is common to the three orders. In particular the doping
dependence reflects the strong anisotropy between hole and electron doping,
which has been connected with the relevance of the $n=5$ Mott insulator on the
whole doping phase diagram of the IBS \cite{werner_prl2008, ishida_prb2010,
nosotras_prb2012-2, demedici_prl2014, nosotras_prb2014, fanfarillo_prb2015}.

\vspace{0.12cm}

\section{Band splittings at symmetry points}

In the unperturbed tetragonal state the $zx$ and $yz$ bands at the $\Gamma$ and
$M$ points are degenerate. The breaking of the $C_4$ symmetry in the nematic
state induces finite splittings $Sp^m_\Gamma$ and $Sp^m_M$ between $zx$ and $yz$
bands at these points which depend on the nematic order and have been
extensively studied in FeSe by ARPES
experiments\cite{ishizaka_prb2014,ishizaka_prb2015,coldea_prb2015,
ding_prb2015,shen_FeSenematic2015,watson2016,brouet2016, borisenko2016}. The
measured ratio $Sp^m_\Gamma/Sp^m_M$ and its relative sign have led to
conclusions on the kind of orbital order present in FeSe. The interpretation of
the experiments has been done assuming the predicted values of the splittings in
the absence of interactions:
\begin{eqnarray}
\nonumber
Sp^{OFO}_\Gamma(U=0)= 2 \delta \epsilon \quad && \quad Sp^{OFO}_M(U=0)=  2 \delta \epsilon
\\
\nonumber
Sp^{SCO}_\Gamma(U=0) = 2\delta t^{\prime} \quad && \quad Sp^{SCO}_M(U=0)= -2\delta t^{\prime}
\\
Sp^{DBO}_\Gamma(U=0) = 0 \phantom{\delta}\quad && \quad  Sp^{DBO}_M(U=0)=  2 \delta t
\label{eq:Splitting0}
\end{eqnarray}  
The renormalization of orbital masses and onsite energies induced by the local
correlations modifies the electronic band structure in the nematic phase.  In
particular, at finite $U$ an onsite orbital-dependent shift $\delta
\lambda_\gamma$ with $\lambda_{zx}\neq \lambda_{yz}$ is induced and, as
discussed above, the quasiparticle weight is renormalized with $Z_{zx}\neq
Z_{yz}$, i.e. the mass is different for $zx$ and $yz$ orbitals. Both effects
renormalize in a moment-dependent way the band structure defining new
$Sp^m_\Gamma$ and $Sp^m_M$, which ratio is qualitatively different with respect
the bare ones of Eq. (\ref{eq:Splitting0}). Even if a quantitive comparison with
experiments\cite{ishizaka_prb2014,ishizaka_prb2015,coldea_prb2015,
ding_prb2015,shen_FeSenematic2015,watson2016,brouet2016, borisenko2016}, at
present controversial, is beyond the aim of this work, in the following we
discuss qualitatively the change in the splittings $Sp^m_\Gamma$ and $Sp^m_M$ 
induced by local correlations. Not to include the effect of this modification
can lead to a wrong interpretation of the experiments.

The primary effect of the correlations is a suppression of the
splittings which follows the band narrowing. This reduction of the splitting
does not modify the ratio $Sp^m_\Gamma/Sp^m_M$ nor it implies an equal
suppression of the nematic order parameters which are described by $\chi_m$ in
Fig.1(a). In particular, in the SCO and DBO orders the effective nematic
hoppings $\delta t'$ and $\delta t$ are approximately renormalized by a factor
$\delta Z_{av} \sim (Z_{zx}+Z_{yz})/2$, being the splittings at $\Gamma$ and $M$
corresponding to these orders suppressed by the same factor. 

However, there is an extra contribution which affects differently $\delta
Sp^m_\Gamma$ and $\delta Sp^m_M$ and changes their ratio. This contribution can
be approximated by\cite{ref_approx}: 
\begin{eqnarray}
\nonumber
\delta Sp^m_\Gamma=-\delta \lambda^m+\delta Z^m (2t^y_{zx,zx}+4t'_{zx,zx})
\\
\delta Sp^m_M=-\delta \lambda^m-\delta Z^m (2t^y_{zx,zx}+4t'_{zx,zx})
\label{eq:approxsplitting}
\end{eqnarray}
$\delta \lambda^m=\lambda^m_{zx}-\lambda^m_{yz}$ modifies the 
onsite splitting and $\delta Z^m=Z_{zx}-Z_{yz}$ induces a momentum-dependent change.

In order to highlight the different evolution of the splittings at $\Gamma$ and
$M$ under the different perturbations we use a numerical value $2h_m = 75 \,
meV$. Fig.~\ref{fig:Fig3}(a) shows the band dispersion along $\Gamma-M$ for the
OFO perturbation and $U \sim 3.5 \, eV$. The splitting at $M$ is much larger
than the one at $\Gamma$ which almost vanishes. This is contrary to the equal
value of the bare splittings $Sp^{OFO}_\Gamma=Sp^{OFO}_M$ of Eq.
(\ref{eq:Splitting0}) assumed in previous works to interpret ARPES experiments. 

The evolution with interactions of $Sp^m_\Gamma$ and $Sp^m_M$ and their ratio
for the three orders is plotted in Figs.~\ref{fig:Fig3}(b-d). A primary effect
of the correlations is to decrease the size of the splittings, as expected from
the renormalization of the bands. Less obvious is the k-dependence of the
renormalization introduced by interactions, that makes the correlated ratios
$Sp^{m}_\Gamma/Sp^{m}_M$, shown in the insets of Figs.~\ref{fig:Fig3}(b-d),
different from the bare ones. This effect is mostly due to the differentiation
of the zx/yz spectral weight $\delta Z^{m} \neq 0$, see
Eq.\ref{eq:approxsplitting}. The splittings $Sp^m_\Gamma$ and $Sp^m_M$ and their
ratio $Sp^{m}_\Gamma/Sp^{m}_M$ are especially sensitive to the interactions at
the Hund's metal crossover. Nevertheless,  $Sp^{m}_\Gamma/Sp^{m}_M$ is affected
differently for the different orders. This ratio is weakly affected in the SCO.
On the contrary in the OFO and DBO cases, the interactions renormalize
differently the band structure at $\Gamma$ and $M$ and can even produce
accidental sign-change between the splittings at the electron and hole pockets,
as for example in Fig.~\ref{fig:Fig3}(b).


\section{Summary and Discussion}

In this work, we have shown that local correlations, and in particular, Hund's
physics, have a strong impact on orbital orders states. The sensitivity is
largest for interaction values at the Hund's metal crossover. The mass
renormalization factors observed experimentally \cite{maletz_prb2014,
coldea_prb2015, ishizaka_prb2015}, place this material at the entrance of the
Hund's metal where we find a more pronounced sensitivity of the nematic
properties to local correlations. At these values of the interactions we find
the largest difference between the quasiparticle weight of $zx$ and $yz$
material, what makes that these two orbitals present different band mass.
For the perturbation used in the splitting analysis, $2 h_m = 75 \,meV$, 
that gives $M$ splittings of the order of $20 meV$ at values of the interactions 
relevant for FeSe ($U\sim 3.5 \,eV$, $J_H/U=0.20$), we found a 
differentiation $\delta Z^{OFO} \sim 0.025$ where the unperturbed $Z_{xz/yz}\sim 0.26$. 
This corresponds to a variation of the $9.5\%$, that reduce to the $3.2\%$ and $1.5 \%$ 
for the SCO and DBO cases respectively. 

While the effect of local interactions may look small on a first sight, it has strong implications on
observable properties, like the splittings $Sp^m_\Gamma$ and $Sp^m_M$ between
the $zx$ and $yz$ bands at $\Gamma$ and $M$ measurable by ARPES. It modifies
their ratio and may even modify their sign. While in the absence of interactions
the splittings at $\Gamma$ and $M$ are expected to be equal for the onsite OFO
order, for interaction values relevant in FeSe the splitting at $\Gamma$ is
strongly reduced with respect to the value at $M$. In the d-wave nematic bond
order DBO, the splitting at $\Gamma$, zero in the absence of interactions,
acquires a finite value approximately 1/3 of the splitting at $M$ and opposite
sign. The relative sign of the splittings is less affected in the case of a
sign-changing SCO order. The modification of the splittings induced by local
correlations has to be taken into account in order to correctly interpret the
ARPES spectra.

From the analysis of the orbital response, we did not find any
evidence of divergence of the orbital nematic susceptibilities $\chi_m$ and
therefore no signature of the associated phase transitions. This means that the
nematic transition  must be either mediated or at least assisted by degrees of
freedom not included here, e.g. spin fluctuations. 
Nevertheless, our study demonstrates that Hund's physics strongly constraints the 
nature of a possible orbital ordered state realized in the nematic phase 
suppressing any order generating large $zx/yz$ charge imbalance as the onsite 
ferro-orbital ordering OFO. On the contrary a sign-changing orbital order SCO, 
with small charge imbalance between $zx$ and $yz$ orbitals, is only weakly 
affected by interactions. Thus SCO emerges as the most likely candidate to be 
realized in the nematic phase of FeSe. This order was proposed to be stabilized, 
by spin fluctuations on the basis of parquet renormalization group 
calculations\cite{chubukov2016}.
In fact, the modification of the splittings at $\Gamma$ and $M$ produced by the 
correlations happens to reduce the difference in occupation between $zx$ and 
$yz$ orbitals. A similar tendency is derived via orbital-selective 
self-energy\cite{brouet2016} and vertex corrections\cite{kontani_prl2016}.

While we have focused on FeSe, our description is not restricted to this
compound. The smaller mass enhancement  measured for iron pnictides with respect
to FeSe suggest that the effect of the Hund's physics on the nematic state of
FeAs compound should be present, even if less pronounced than in FeSe.\\

Note added: After we submitted this article for the first time the
preprint\cite{kreisel2016} was posted on the arXiv. In this work the
Quasiparticle Interference spectrum is fitted using a multi-orbital RPA approach
which uses the quasiparticle weights as fitting parameters. Interestingly
different values for $Z_{zx}$ and $Z_{yz}$ are allowed. However, a very large
difference between $Z_{zx}~0.16$ and $Z_{yz}~0.8$ is obtained from these
fittings. Such a large difference cannot accounted for by local correlations. 
In fact if such difference between the mass of these orbitals were associated to
a renormalization of the hopping parameters the electronic bandstructure would
dramatically modified and this is not observed experimentally.

\section*{Acknowledgments}
We thank conversations with M.J. Calder\'on, A. Chubukov, A. Coldea, R.
Fernandes, L. Rhodes, R. Valenti, B. Valenzuela and M.D. Watson. L.F. thanks
A.Valli for his invaluable help during the data analysis. E.B. acknowledges
funding from Ministerio de Econom\'ia y Competitividad via grants No.
FIS2011-29689 and FIS2014-53218-P and from Fundaci\'on Ram\'on Areces. M.C.
acknowledges funding by SISSA/CNR project "Superconductivity, Ferroelectricity
and Magnetism in bad metals" (Prot. 232/2015) and MIUR through the PRIN 2015
program (Prot. 2015C5SEJJ001).
\bibliography{nematic}

\begin{thebibliography}{63}
\bibitem[{ref({\natexlab{a}})}]{ref_plane}
\bibinfo{note}{The average values used to calculate $\chi_{SCO}$ and
  $\chi_{DBO}$ in Fig. 1(a) and the splittings $Sp^m_\Gamma$ and $Sp^m_M$ and
  band structure in Fig. 3 have been computed neglecting the $k_z$ dependence
  of the band structure. This approximation do not alter our conclusions.}
\bibitem[{hop()}]{hopping_note}
\bibinfo{note}{We have found similar conclusions to that found for the SCO for
  a perturbation of the kind $\sim n_{zx}\cos ky - n_{yz}\cos kx$ which also
  changes sign between electron and hole pockets. This perturbation is a
  combination of $A_{1g}$ and $B_{1g}$ symmetries which breaks the degeneracy
  of the largest intraorbital $zx$ and $yz$ nearest neighbor hoppings.}
\bibitem[{ref({\natexlab{b}})}]{ref_OFO}
\bibinfo{note}{$2\epsilon^*_{OFO}$ decreases with interactions faster than the
  quasiparticle weights of zx and yz orbitals. This faster decrease is
  responsible for the suppression of the occupation imbalance.}
\bibitem[{ref({\natexlab{c}})}]{ref_approx}
\bibinfo{note}{Only the largest hopping terms in the tight-binding have been
  included in the $\delta Z^m$ term.}
\end{thebibliography}
\end{document}